# Observation of Low-frequency Interlayer Breathing Modes in Few-layer Black Phosphorus

*Xi Ling*[1,*], *Liangbo Liang*[2,*], *Shengxi Huang*[1], *Alexander A. Puretzky*[3], *David B. Geohegan*[3], *Bobby G. Sumpter*[3,4], *Jing Kong*[1], *Vincent Meunier*[2], *Mildred S. Dresselhaus*[1,5]

As a new two-dimensional layered material, black phosphorus (BP) is a promising material for nanoelectronics and nano-optoelectronics. We use Raman spectroscopy and first-principles theory to report our findings related to low-frequency (LF) interlayer breathing modes (<100 $cm^{-1}$) in few-layer BP for the first time. The breathing modes are assigned to $A_g$ symmetry by the laser polarization dependence study and group theory analysis. Compared to the high-frequency (HF) Raman modes, the LF breathing modes are much more sensitive to interlayer coupling and thus their frequencies show much stronger dependence on the number of layers. Hence, they could be used as effective means to probe both the crystalline orientation and thickness for few-layer BP. Furthermore, the temperature dependence study shows that the breathing modes have a harmonic behavior, in contrast to HF Raman modes which are known to exhibit anharmonicity.

---

[1]Department of Electrical Engineering and Computer Science, Massachusetts Institute of Technology, Cambridge, Massachusetts 02139, USA. [2]Department of Physics, Applied Physics, and Astronomy, Rensselaer Polytechnic Institute, Troy, New York 12180, USA. [3]Center for Nanophase Materials Sciences, Oak Ridge National Laboratory, Oak Ridge, Tennessee 37831, USA. [4]Computer Science and Mathematics Division, Oak Ridge National Laboratory, Oak Ridge, Tennessee 37831, USA. [5]Department of Physics, Massachusetts Institute of Technology, Cambridge, Massachusetts 02139, USA. *These authors contributed equally to this work. Correspondence should be addressed to M.S.D. (email: mdress@mit.edu), V.M. (email: meuniv@rpi.edu) and X.L. (email: xiling@mit.edu).

Orthorhombic black phosphorus (BP) is the most stable allotrope of phosphorus. It features a layered structure with puckered monolayers stacked by van der Waals (vdW) force[1]. Few- or single-layer BP can be mechanically exfoliated from bulk BP[2–5]. Due to BP's intrinsic thickness-dependent direct bandgap (ranging from 0.3 eV to 2.0 eV) and relatively high carrier mobility (up to ~1,000 $cm^2 V^{-1} s^{-1}$ at room temperature)[2,4,6–9], it is expected to have promising applications in nanoelectronic devices[2–4,10], and near and mid-infrared photodetector[11–19]. Recently, high performance thermoelectric devices were also predicted based on BP thin films[20–22]. With the surge of interest in two-dimensional (2D) materials (such as graphene and transition metal dichalcogenides (TMDs))[23,24], BP has become a new attraction since 2014 because it bridges the gap between graphene and TMDs, and offers the best trade-off between mobility and on-off ratio[3]. Moreover, the unique anisotropic puckered honeycomb lattice of BP leads to many novel in-plane anisotropic properties, which could lead to more applications based on BP[3,9,25,26].

Phonon behaviors play an important role in the diverse properties of materials[27], which has been intensively studied in vdW layered materials, such as graphene and TMDs[28–35]. Raman spectroscopy is a powerful non-destructive tool to investigate the phonons and their coupling to electrons, and has been successfully applied to vdW layered materials[36–40]. Due to the lattice dynamics of vdW layered compounds, the phonon modes in these materials can be classified as high-frequency (HF) intralayer modes and low-frequency (LF) interlayer modes[27]. Intralayer modes involve vibrations from the intralayer chemical bonds (Fig. 1c), and the associated frequencies reflect the strength of those bonds. In contrast, the interlayer modes correspond to layer-layer vibrations with each layer vibrating as a whole unit (Fig. 1b), and hence their frequencies are determined by the interlayer vdW restoring forces. The weak nature of vdW interactions renders the frequencies of interlayer modes typically much lower than those of intralayer modes, usually below 100 cm$^{-1}$. Depending on the vibrational direction, LF interlayer modes are categorized into two types: the in-plane shear modes and the out-of-plane breathing mode (Fig. 1b). Compared to the HF intralayer modes, they are more sensitive to both the interlayer coupling and thickness. The LF interlayer modes have been shown to be very important in studying the interlayer coupling and identifying the thickness for few-layer graphite and TMDs[41–43].

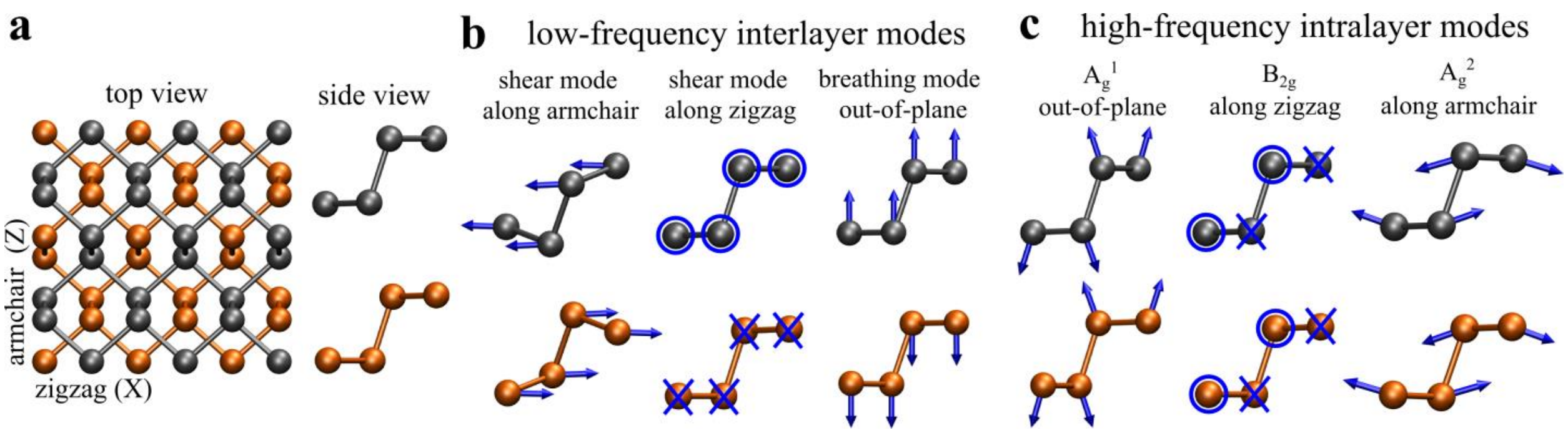

**Figure 1. Crystal structure and phonon vibrations of orthorhombic black phosphorus (BP).** (a) Top and side views of BP with puckered layers. The top and bottom layers are differentiated using black and gold colors. (b) Vibrations of LF interlayer modes: two in-plane shear modes and one out-of-plane breathing mode. (c) Vibrations of HF intralayer modes: three characteristic Raman modes $A_g^1$, $B_{2g}$ and $A_g^2$. The circle and cross indicate vibrations coming out of the page or going into the page.

The HF intralayer Raman modes in bulk BP crystals were studied in the 1980s[44] and recently similar HF modes have been reported in thin film BP[3,45–47]. Normally, three characteristic HF Raman modes ($A_g^1$, $B_{2g}$ and $A_g^2$) can be observed under the typical back-scattering configuration, corresponding to the out-of-plane vibration (~365 cm$^{-1}$), in-plane vibrations along the zigzag direction (~440 cm$^{-1}$) and armchair direction (~470 cm$^{-1}$), respectively (Fig. 1c). Moreover, it is found that their excitation laser polarization dependence can be used to determine the crystal orientation of BP[45,47]. However, the frequencies of HF intralayer modes are found to exhibit almost no dependence on film thickness[44–47]. Clearly, the study of the LF interlayer phonon modes in few-layer BP (Fig. 1b) is needed to reveal more information on the interlayer coupling and thickness. They have been studied in bulk BP using inelastic neutron scattering in the1980s[48,49]. However, to the best of our knowledge, there has been no experimental work on the observation of LF interlayer modes in few-layer BP, except two recent theoretical works[50,51]. The measurement of LF (<100 $cm^{-1}$) Raman modes is challenging since these modes are usually blocked by the Rayleigh rejecter, and it requires a Raman system with LF coupler or triple-grating Raman system. In this work, we successfully observed the LF interlayer breathing Raman modes in few-layer BP for the first time. These breathing modes are assigned to $A_g$ symmetry based on an experimental laser polarization dependence analysis and first-principles density functional theory (DFT) calculations. The thickness dependence study indicates that the breathing modes in few-layer BP are strongly thickness-dependent, and thus could be used as an important and effective indicator of the number of layers. In addition, the temperature dependence study indicates that the interlayer breathing modes present harmonic phonon effects, while the HF intralayer modes show anharmonic phonon behaviors.

## Results

### Theoretical prediction and experimental observation of the interlayer breathing modes

According to our symmetry analysis[52,53], bulk BP crystals belong to the space group *Cmce* (No. 64) and point group $D_{2h}^{18}$ (mmm)[44]. As shown in Fig. 1a, the crystal unit cell of bulk BP is orthorhombic with two layers and 8 atoms (a=~3.3, b=~10.5 and b=~4.4 Å). The primitive unit cell is half of the crystal unit cell and contains 4 atoms, and hence there are 12 normal phonon modes at the Γ point:

$\Gamma_{bulk} = 2A_g + B_{1g} + B_{2g} + 2B_{3g} + A_u + 2B_{1u} + 2B_{2u} + B_{3u}$, (1)

where $A_g$, $B_{1g}$, $B_{2g}$, $B_{3g}$ modes are Raman-active, $B_{1u}$, $B_{2u}$, $B_{3u}$ modes are infrared-active, and $A_u$ mode is optically inactive[44,45,54]. According to the classical Placzek approximation[55], Raman intensity of a phonon mode is proportional to $|e_i \cdot \tilde{R} \cdot e_s^T|^2$, where $e_i$ and $e_s$ are the electric polarization vectors of the incident and scattered light respectively, and $\tilde{R}$ is the Raman tensor of the phonon mode. Only when $|e_i \cdot \tilde{R} \cdot e_s^T|^2$ is not zero, can the phonon mode be observed by Raman spectroscopy. As a common practice in the literature[43,44,52], we denote the in-plane zigzag direction as X axis, the out-of-plane direction as Y axis, and in-plane armchair direction as Z axis. The calculated Raman tensors $\tilde{R}$ of Raman-active modes $A_g$, $B_{1g}$, $B_{2g}$ and $B_{3g}$ are

$$\tilde{R}(A_g)=\begin{pmatrix} a & \cdot & \cdot \\ \cdot & b & \cdot \\ \cdot & \cdot & c \end{pmatrix}, \qquad \tilde{R}(B_{1g})=\begin{pmatrix} \cdot & d & \cdot \\ d & \cdot & \cdot \\ \cdot & \cdot & \cdot \end{pmatrix},$$

$$\tilde{R}(B_{2g})=\begin{pmatrix} \cdot & \cdot & e \\ \cdot & \cdot & \cdot \\ e & \cdot & \cdot \end{pmatrix}, \qquad \tilde{R}(B_{3g})=\begin{pmatrix} \cdot & \cdot & \cdot \\ \cdot & \cdot & f \\ \cdot & f & \cdot \end{pmatrix}, \tag{2}$$

where *a*-*f* are major terms while other terms (denoted by "·") are either zero or negligible due to symmetry[45,56,57]. In the typical experimental back-scattering laser configuration, the electric polarization vectors $e_i$ and $e_s$ are in-plane (the X-Z plane), and thus only $A_g$ and $B_{2g}$ modes can be observed according to the Raman tensors, although $B_{1g}$ and $B_{3g}$ are Raman-active (more details in Supplementary information (SI))[42,45–47]. The symmetries of *N*L BP films (where *N*L is the number of layers) are slightly different from those of bulk BP: odd *N*L BP belong to the space group *Pmna* (No. 53) and the point group $D_{2h}^{7}$ (mmm); even *N*L BP belong to the space group *Pmca* (No. 57) and the point group $D_{2h}^{11}$ (mmm). Although *N*L systems belong to different space groups from the bulk BP, all of them share the same point group $D_{2h}$(mmm). Consequently, the symmetry classification of Raman modes and the forms of their Raman tensors remain unchanged for any thickness (Eq.1 and Eq. 2), consistent with previous theoretical works[51,58].

In $N$L BP, there are $N-1$ interlayer shear modes vibrating along the zigzag direction, $N-1$ interlayer shear modes along the armchair direction, and $N-1$ interlayer breathing modes along the out-of-plane direction, similar to 2D graphene and TMDs[41,43,50]. The difference is that the shear modes vibrating along zigzag and armchair directions are non-degenerate in BP due to its in-plane anisotropy. For perfect (defect-free and free-standing) BP films, the shear modes are either Raman-active ($B_{1g}$ or $B_{3g}$) or infrared-active ($B_{1u}$ or $B_{3u}$), while the breathing modes are either Raman-active ($A_g$) or infrared-active ($B_{2u}$)[51]. As discussed in Eq. 2, in the back-scattering configuration, only $A_g$ and $B_{2g}$ modes can be detected by Raman spectroscopy. Consequently, among the LF interlayer modes, only the Raman-active breathing modes with $A_g$ symmetry can be observed in our Raman spectra. Furthermore, the number of breathing modes with Raman-active $A_g$ symmetry is $N/2$ for even $N$L and $(N-1)/2$ for odd $N$L (see Table 1)[51]. For monolayer BP (or phosphorene), the interlayer breathing modes do not exist. Bulk BP has a breathing mode (around 87 $cm^{-1}$)[48–50,54], but its calculated Raman tensor $\tilde{R}$ is zero, indicating that it cannot be detected. Therefore, in short, the breathing modes can only be observed in few-layer BP, not in single-layer and bulk BP. In addition, according to the inelastic neutron scattering measurements on bulk BP[48–50,54], the two shear modes (vibrating along armchair and zigzag directions respectively) have frequencies around 19 and 52 $cm^{-1}$, while the frequency is ~87 $cm^{-1}$ for the breathing mode. From our calculations and previous theoretical works[50,51], in few-layer BP, the frequencies of all shear modes are no larger than their bulk values (thus $\leq 52$ $cm^{-1}$); similarly the frequencies of all breathing modes are no larger than their bulk values (thus $\leq 87$ $cm^{-1}$). These results suggest that LF peaks observed above 52 $cm^{-1}$ are likely to be breathing modes.

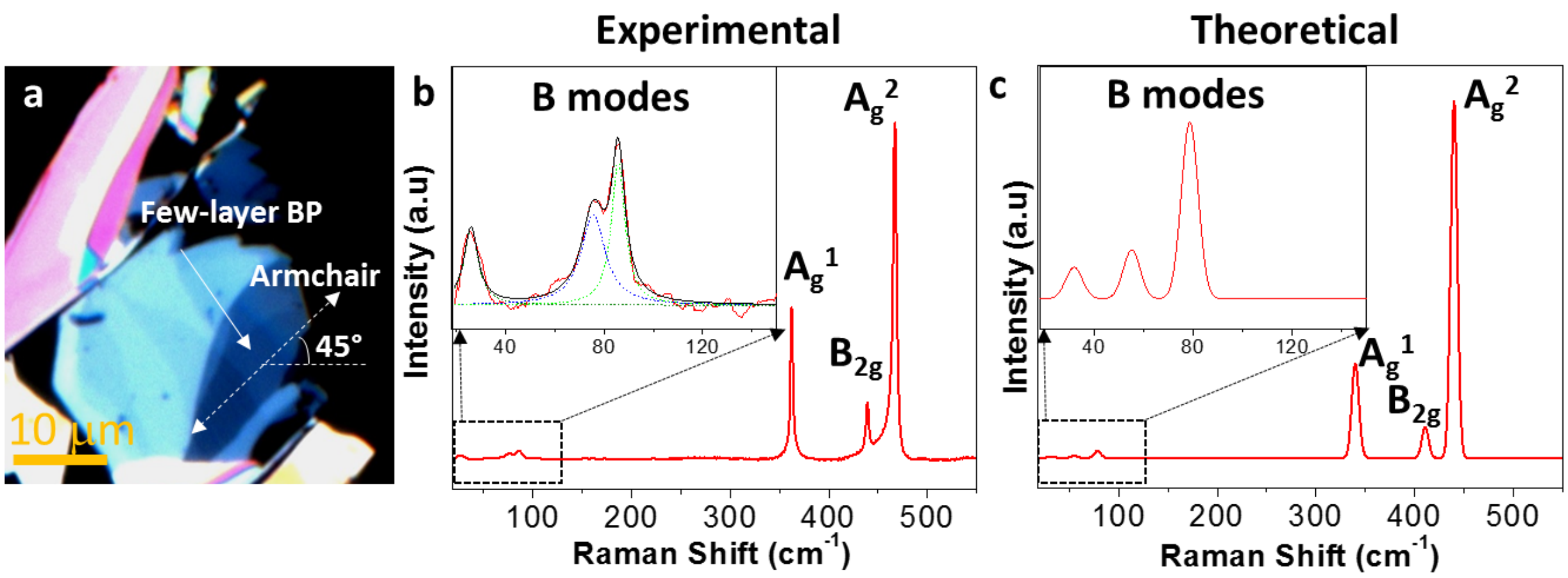

**Figure 2. Low-frequency Raman modes in few-layer BP.** (a) A typical optical image of exfoliated BP flakes on a glass substrate, including few-layer BP (the blue area). (b) Experimental Raman spectrum of few-layer BP corresponding to the flake in (a). Inset: the zoom-in spectrum from 20 to 150 $cm^{-1}$. (c) Calculated Raman spectrum of 6L BP in the experimental back-scattering geometry. Inset: the zoom-in spectrum in the same LF region as (b). Three interlayer breathing modes (B modes) with Raman-active $A_g$ symmetry are predicted in the LF region.

The experimental Raman measurements were carried out on few-layer BP flakes (Fig. 2). The BP flakes on a glass substrate were mechanically exfoliated from the bulk and coated with parylene (~100 nm) or PMMA film (~300 nm) immediately to avoid degradation. From the optical contrast of the flakes, the bluish flakes are determined as few-layer BP, while the reddish and whitish flakes are thicker ones[2,3,59]. The corresponding Raman spectrum on the few-layer BP (the blue area labeled in Fig. 2a) is shown in Fig. 2b. The three well-known HF $A_g^1$, $B_{2g}$ and $A_g^2$ peaks of BP are located at 362.3 $cm^{-1}$, 439.2 $cm^{-1}$, and 467.1 $cm^{-1}$, respectively. More interestingly, another three peaks with relatively weaker intensities are observed in the LF region. As shown in the zoom-in spectrum in the inset of Fig. 2b, the frequencies of these three peaks are determined by peak fitting as 26.2 $cm^{-1}$, 75.6 $cm^{-1}$ and 85.6 $cm^{-1}$, respectively.

According to our theoretical analysis, they are expected to be LF interlayer breathing modes (labeled as "B modes") belonging to Raman-active $A_g$ symmetry. In addition, only when $N \geq 6$, can there be no less than three B modes with $A_g$ symmetry. Therefore, we conclude that the number of layers of the measured few-layer BP flake in Fig. 2a is at least 6. The calculated Raman spectrum of 6L BP is shown in Fig. 2c. Besides the $A_g^1$, $B_{2g}$ and $A_g^2$ modes in the HF region, three B modes appear in the LF region with their frequencies located around 31.9 $cm^{-1}$, 55.1 $cm^{-1}$ and 78.6 $cm^{-1}$ (inset in Fig. 2c), confirming our interpretation of the experimental observations. To provide further experimental evidence that the three LF Raman peaks in Fig. 2b are B modes with $A_g$ symmetry, we performed laser polarization dependence measurements of all Raman modes, as shown in the following section.

**Polarization dependence**

Due to the in-plane anisotropic structure of the BP thin film, its Raman modes show significant polarization dependence, which have been demonstrated to be used to identify the crystal orientation of the sample[45,47]. Two methods were reported to study the polarization dependence. One is by rotating the sample while fixing the polarization of the incident and scattered light[45]; the other is by changing the polarization of the incident and scattered light while fixing the sample[47]. Here we used the first method. As discussed above, the Raman intensity is $I \propto |e_i \cdot \tilde{R} \cdot e_s^T|^2$. In the experimental back-scattering geometry, the electric polarization vectors $e_i$ and $e_s$ of the incident and scattered light are in-plane (the X-Z plane: X (Z) axis is defined as sample initial zigzag (armchair) direction before rotating the sample). By setting the polarization angle of the incident (scattered) light as $\theta$ ($\gamma$) with respect to X axis, we have $I \propto \left|(cos\theta, 0, sin\theta) \quad \tilde{R} \quad \begin{pmatrix} cos\gamma \\ 0 \\ sin\gamma \end{pmatrix}\right|^2$. For the sample rotation method, $\theta$ and $\gamma$ are fixed, and the

sample is rotated in-plane (the X-Z plane) by $\varphi$ with respect to X axis. The Raman intensity then becomes:

$$I \propto \left| (\cos(\theta-\varphi), 0, \sin(\theta-\varphi)) \quad \tilde{R} \quad \begin{pmatrix} \cos(\gamma-\varphi) \\ 0 \\ \sin(\gamma-\varphi) \end{pmatrix} \right|^2$$ (more details in SI).

In our experiment, we used the parallel polarization configuration, so that $\gamma = \theta$ always. For an $A_g$ mode, its Raman tensor is $\tilde{R} = \begin{pmatrix} a & \cdot & \cdot \\ \cdot & b & \cdot \\ \cdot & \cdot & c \end{pmatrix}$, thus

$$I_{A_g} \propto a^2 \left| 1 + (\frac{c}{a} - 1) sin^2(\varphi - \theta) \right|^2 . \quad (3)$$

Since $\theta$ is fixed, the intensity of an $A_g$ mode depends on both the sample rotation angle $\varphi$ and the ratio $c/a$. For a $B_{2g}$ mode, the Raman tensor $\tilde{R} = \begin{pmatrix} \cdot & \cdot & e \\ \cdot & \cdot & \cdot \\ e & \cdot & \cdot \end{pmatrix}$, thus

$$I_{B_{2g}} \propto e^2 sin^2 2(\varphi - \theta), \quad (4)$$

which only depends on the rotation angle $\varphi$, since $\theta$ is fixed. According to our calculations and a previous experimental work[45], $c$ (tensor component related to the armchair direction) is expected to be larger than $a$ (tensor component related to the zigzag direction) in Eq. 3, hence $c/a > 1$. Therefore, the minimum intensity angle of an $A_g$ mode is $\varphi = \theta$ or $\theta + 180°$ (the sample zigzag direction is now rotated to the polarization direction of incident light); the maximum intensity angle of an $A_g$ mode is $\varphi = \theta + 90°$ or $\theta + 270°$ (the sample armchair direction is now rotated to the polarization direction of incident light). For a $B_{2g}$ mode, its minimum intensity angle is $\varphi = \theta$ or $\theta + 90°$ or $\theta + 180°$ or $\theta + 270°$ (the sample armchair or zigzag direction is now rotated to the polarization direction of incident light); its maximum intensity angle is $\varphi = \theta + 45°$ or $\theta + 135°$ or $\theta + 225°$ or $\theta + 315°$. Hence, by rotating the sample under parallel polarization configuration, the intensity variation period is always 180° for an $A_g$ mode, while it is 90° for a $B_{2g}$ mode. In addition, when the sample armchair (zigzag) direction is along the polarization direction of incident light, an $A_g$ mode shows the maximum (minimum) intensity, while a $B_{2g}$ mode is forbidden[45]. These results can be used to identify the crystalline orientation.

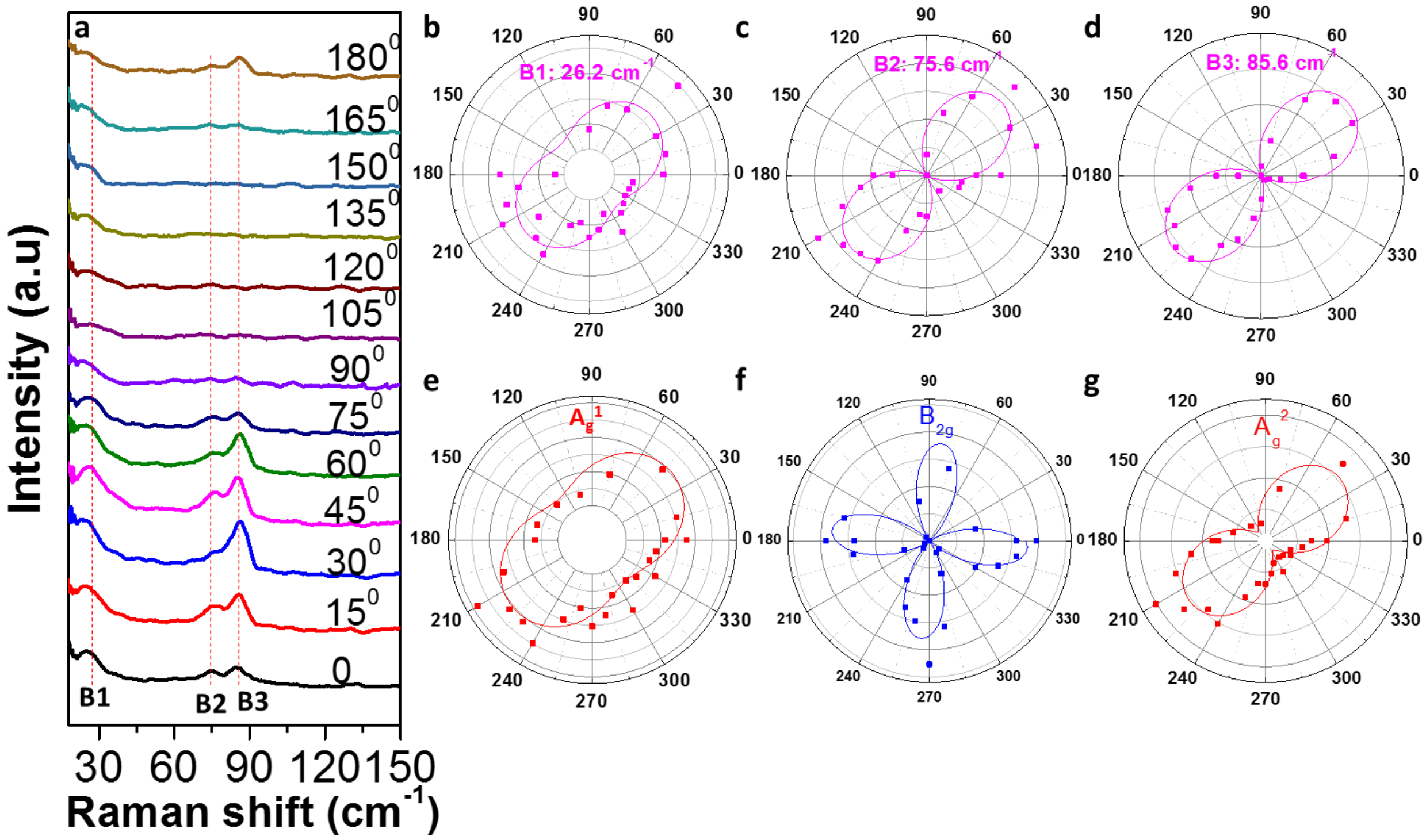

**Figure 3. Polarization dependence of both LF and HF modes.** (a) LF Raman spectra of few-layer BP (corresponding to Fig. 2a) at different sample rotation angles. (b-d) The profile of the intensities of the LF Raman modes at different rotation angles. (b): B1: 26.2 cm$^{-1}$; (c) B2: 75.6 cm$^{-1}$; (d) B3: 85.6 cm$^{-1}$. (e-f) The profile of the intensities of the HF Raman modes at different rotation angles. (e): $A_g^1$; (f) $B_{2g}$; (d) $A_g^2$. The sample was rotated clockwise from 0° to 360°.

Fig. 3a shows a series of Raman spectra of few-layer BP in the LF region at different sample rotation angles. The corresponding Raman spectra in the HF region are shown in Fig. S3. With the sample was rotated from 0° to 180°, the intensities of the three LF modes vary periodically, and reach the maximum around 45° and the minimum around 135°. It should be mentioned that we could not differentiate the LF modes from the background noise for the rotation from 105° to 165° because they are too weak at these polarization values. These results clearly establish the importance of carefully considering the rotation of the sample when studying the LF modes of BP. The polar plots of the fitted peak intensities of both the LF and HF modes as a function of the rotation angle are shown in Figs. 3b-g. The three LF modes (Figs. 3b-d) and HF $A_g^1$ and $A_g^2$ modes (Figs. 3e and 3g) share very similar polarization dependence: all of them have the same intensity variation period of 180° with two intensity maxima around 45° or 225°. However, the HF $B_{2g}$ mode shows the intensity variation period of 90° with four intensity maxima around 0°, 90°, 180°, 270° (Fig. 3f). These are consistent with our theoretical predictions above and the theoretical polar plot in Fig. S2. The polarization dependence measurement provides another piece of evidence that the three LF modes share the same symmetry as the HF $A_g^1$ and $A_g^2$ modes (i.e., $A_g$ symmetry). These three LF modes are thus assigned to interlayer breathing modes that have $A_g$ symmetry, since shear modes (belonging to $B_{1g}$ or $B_{3g}$ symmetry)[51] have different polarization dependence from the $A_g$ modes. The polarization dependence study also shows that the armchair direction of the sample is about the 45° direction of the image we presented in Fig. 2a (more details in SI). Note that although LF breathing (B) modes and HF $A_g$ modes share very similar polarization dependence in Figs. 3b-g, there are still minor differences. At the minimum intensity rotation angle (~135° or 315°), one should note that the LF B1 and HF $A_g^1$ and $A_g^2$ modes show relatively strong intensities, while the LF B2 and B3 modes are barely present. This is due to the difference of the *c*/*a* ratio in the Raman tensors of B and $A_g$ modes despite the same symmetry (see Eq. 3 and Fig. S2a).

**Thickness dependence of LF Raman spectra**

As suggested by the polarization dependence, the intensities of the Raman modes of BP are strongly related to the crystal orientation of the sample. Therefore, when studying the thickness dependence of the Raman modes, it is important to set the flakes along the same crystal orientation. It is not rigorous to discuss the thickness dependence of the Raman spectra without considering the sample rotation. Here, for every flake chosen for a thickness dependence study, we collected the Raman spectra of the flakes at different orientations and determined the armchair direction of the flakes. The B modes of the different flakes for comparison are all collected with the laser polarization along the armchair direction, at which intensities are the maximum. The optical images and the corresponding Raman spectra of the flakes on 300 nm $SiO_2$/Si substrates with PMMA coating are shown in Fig. 4. Since the sample is polymer coated immediately after exfoliation to avoid degradation, it becomes very difficult to directly measure the thicknesses of the flakes. But from the optical contrast of the flakes, its thickness ordering can be estimated. The flakes get thicker from flake 1 to flake 5. On flake 1, we did not observe any LF peak at any polarization direction (Fig. 4b). In addition, the Raman intensities of the HF modes on flake 1 are very weak (Fig. S4 in SI). These results indicate that flake 1 might be monolayer (recall that monolayer cannot have LF interlayer modes)[12,46,60]. For the few-layer BP in Fig. 4b, from flake 2 to flake 3, a LF B mode appears and the peak splits into two from flake 4 to flake 5.  However, for very thick multilayer (ML) flakes and bulk sample, there is no LF B mode observed (Fig. 4b). The zero intensity of LF modes in bulk BP is consistent with the theoretical analysis outlined above. For very thick ML flakes, they are bulk-like and thus the intensities of LF modes are too low to be detected as well. Only in the few-layer samples (flakes 2-5 in Fig. 4b), LF modes show observable intensities. Such tendency is consistent with other vdW layered materials such as TMDs, where bulk-inactive vibrational modes become Raman-active and observable in few-layer but they are non-detectable in very thick samples[41,61,62].

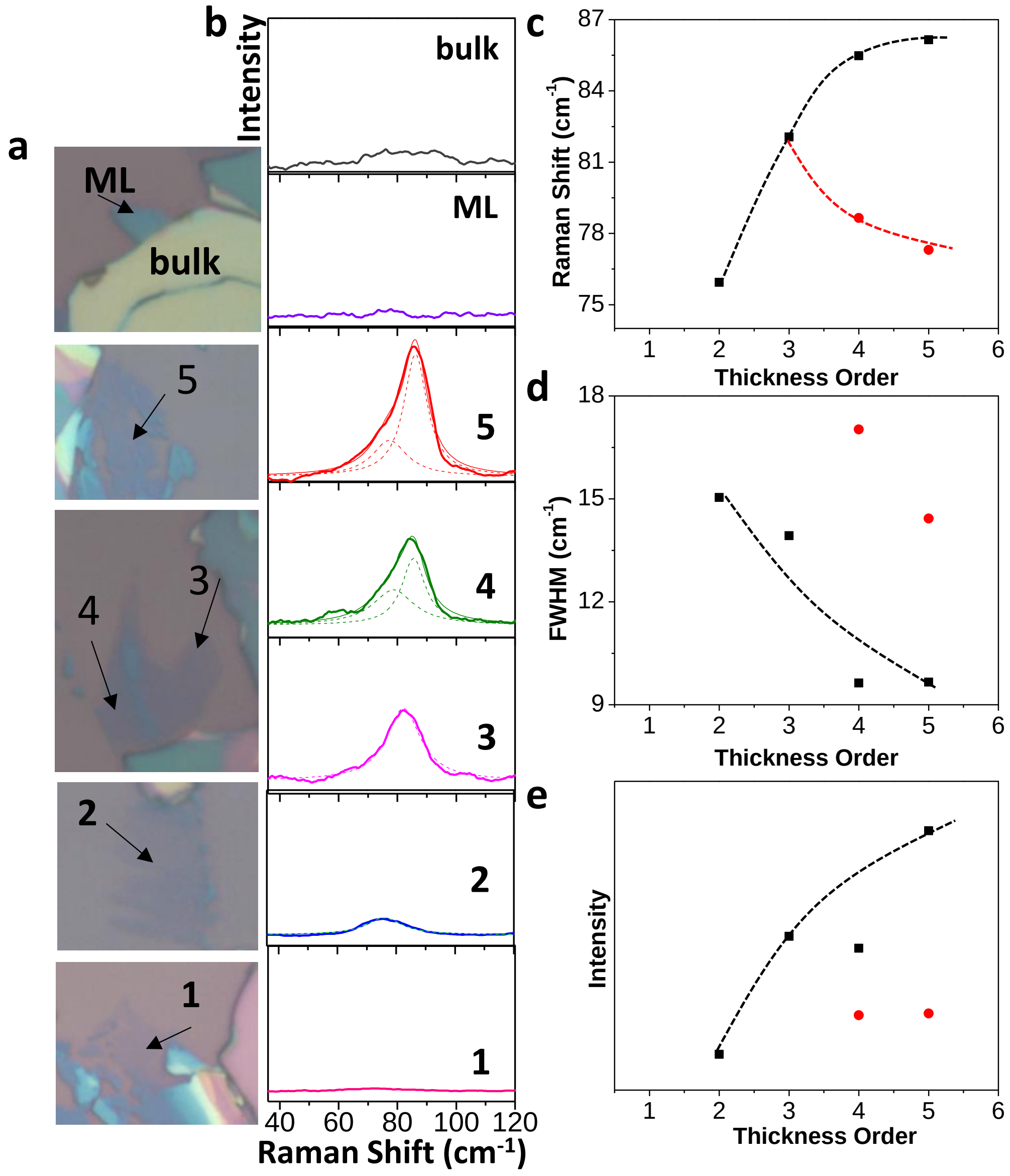

**Figure 4. Thickness dependence of LF breathing (B) modes.** (a) Optical images of BP flakes with different thicknesses. The thickness increases with the flake from bottom to top. (b) LF Raman spectra collected on the flakes corresponding to (a). For flakes 4 and 5, the peak splits into two. (c) Raman shift, (d) FWHM and (e) intensities of the B modes as a function of the thickness. The black (red) points correspond to the higher-frequency (lower-frequency) B mode.

To understand the thickness dependence of B modes, we further calculated the frequencies of B modes of 2L to 8L BP using DFT PBE+optB88 method (Table 1). Such method gives the bulk B mode's frequency 86.1 $cm^{-1}$, very close to the experimental value (87 $cm^{-1}$)[48–50,54]. Other calculation methods have been also used for comparison (more details in Method and Table S1 in SI). In each column of Table 1, the frequency of the B mode monotonically decreases with increasing thickness, well consistent with previous theoretical works on BP[50,51] and experimental reports on TMDs[40,41], as this constitutes a general trend for vdW layered materials. Furthermore, the highest-frequency B mode of any thickness in Table 1 is the bulk-like B mode, where each adjacent layer vibrating in the opposite directions (see the vibrations in Fig. S1). With increasing thickness, it blue shifts and approaches the bulk limit 87 $cm^{-1}$, very similar to the observed higher-frequency mode (the black points in Fig. 4c). It follows that the observed higher-frequency mode in a BP flake should correspond to the flake's highest-frequency B mode (i.e., bulk-like B mode). As for the lower-frequency B mode (the red points in Fig. 4c), it probably corresponds to the second-highest B mode of the flake. Since the highest-frequency B mode is not Raman-active ($B_{2u}$) for odd *N*, the BP flakes 2-5 showing bulk-like B modes might be all even *N*L. Another possibility is that the polymer capping or the supporting substrates or defects in the material may break the symmetry to induce Raman-activation of the bulk-like B modes in odd *N*L BP. A definite conclusion cannot be drawn for now because direct measurements of the flakes' thickness are not possible due to the instability of BP and the polymer protection capping. Nevertheless, regardless of the exact thickness, a major finding in this work is that the frequency changes of the LF B modes in Fig. 4c can be more than 10 $cm^{-1}$, while the frequency variations of HF $A_g^1$, $B_{2g}$ and $A_g^2$ modes with the thickness are much smaller (~2 $cm^{-1}$, see Fig. S4)[44–47,60]. Consequently, the LF modes could offer a more effective approach to determine the thickness and probe the interlayer vdW coupling of BP. Furthermore, the number of Raman-active B modes is zero in 1L, one for 2L and 3L, and more for thicker BP (Fig. 4 and Table 1). Hence, the number of LF peaks can also help to quickly estimate the number of layers. We expect that the present work can stimulate further experimental efforts to identify the thickness and probe the LF modes, thus establishing more conclusive relationship between them.

**Table 1.** Calculated frequencies (in $cm^{-1}$) of breathing modes of 2L to 8L BP using PBE+optB88 method. In *N*L BP, there are *N*−1 breathing modes either Raman-active ($A_g$) or infrared-active ($B_{2u}$). The number of breathing modes with Raman-active $A_g$ symmetry is *N*/2 for even *N*, and (*N*-1)/2 for odd *N*. The breathing modes are labeled as B*n*, and the ones belong to $A_g$ symmetry are highlighted in red color. The B mode of bulk BP is also shown but it cannot be detected.

| Layer number | B1 | B2 | B3 | B4 | B5 | B6 | B7 |
|---|---|---|---|---|---|---|---|
| 2L | 62.7 ($A_g$) | | | | | | |
| 3L | 52.0 ($A_g$) | 70.5 | | | | | |
| 4L | 36.2 ($A_g$) | 63.1 | 75.6 ($A_g$) | | | | |
| 5L | 33.7 ($A_g$) | 53.4 | 69.5 ($A_g$) | 76.2 | | | |
| 6L | 31.9 ($A_g$) | 42.1 | 55.1 ($A_g$) | 71.0 | 78.6 ($A_g$) | | |
| 7L | 28.2 ($A_g$) | 35.7 | 51.7 ($A_g$) | 65.2 | 74.5 ($A_g$) | 80.4 | |
| 8L | 24.8 ($A_g$) | 31.0 | 47.8 ($A_g$) | 60.9 | 71.4 ($A_g$) | 77.6 | 83.2 ($A_g$) |
| bulk | | | | | | | 86.1 |

In addition, we also show the dependence of FWHM (Fig. 4d) and intensity (Fig. 4e) of the B modes. The decrease of the FWHM with the increase of the thickness indicates that the lifetime of the B mode phonons is longer in the thicker flakes, similar to TMDs[63]. In Fig. 4e, the intensity of the higher-frequency mode generally increases with the thickness from flake 2 to 5.

**Temperature dependence**

The temperature dependence of Raman spectra is important for understanding the fine structure and properties of the material, probing phonons and their interactions with other particles, which in turn has a large impact on the electronic and thermoelectric device performances. The temperature dependence of the B mode, $A_g^1$, $B_{2g}$ and $A_g^2$ modes in BP is measured under 632 nm laser excitation from -150 to 30 °C (Fig. 5 and Fig. S5). The data are fitted linearly using the equation[64]: $\omega = \omega_0 + \chi T$ (red lines in Fig. 5), where $\omega_0$ is the frequency at $T$=0 °C and $\chi$ is the first-order temperature coefficient, which defines the slope of the dependence. It can be clearly seen that the temperature dependence behaviors for the different modes are different. In particular, the B mode shows a very weak temperature dependence, which has almost no frequency change in the examined temperature range (Fig. 5a) with $\omega_B = 87.4 - 1.8 \times 10^{-4} T$. This suggests the harmonic property of the B mode in few-layer BP[64,65]. However, anharmonic phonon effect occurs for the HF modes according to the stronger temperature dependence (Figs. 5b-d), where $\omega_{\mathrm{Ag1}} = 361 - 0.0073T$ for $A_g^1$, $\omega_{\mathrm{B2g}} = 438 - 0.013T$ for $B_{2g}$ and $\omega_{\mathrm{Ag2}} = 465 - 0.012T$ for $A_g^2$. Furthermore, the first-order temperature coefficient is larger for the in-plane vibrational modes ($B_{2g}$ and $A_g^2$) than for the out-of-plane vibrational mode ($A_g^1$). This is consistent with the results obtained on the bulk BP[56]. The temperature coefficients of the in-plane Raman modes of few-layer BP (-0.013 $cm^{-1}$/K for $B_{2g}$ mode and -0.012 $cm^{-1}$/K for $A_g^2$ mode) are similar to some other layered materials, such as graphene (-0.015 $cm^{-1}$/K for G band)[65,66] and $MoS_2$ (-0.013 $cm^{-1}$/K for $E_{2g}$ mode)[67,68].

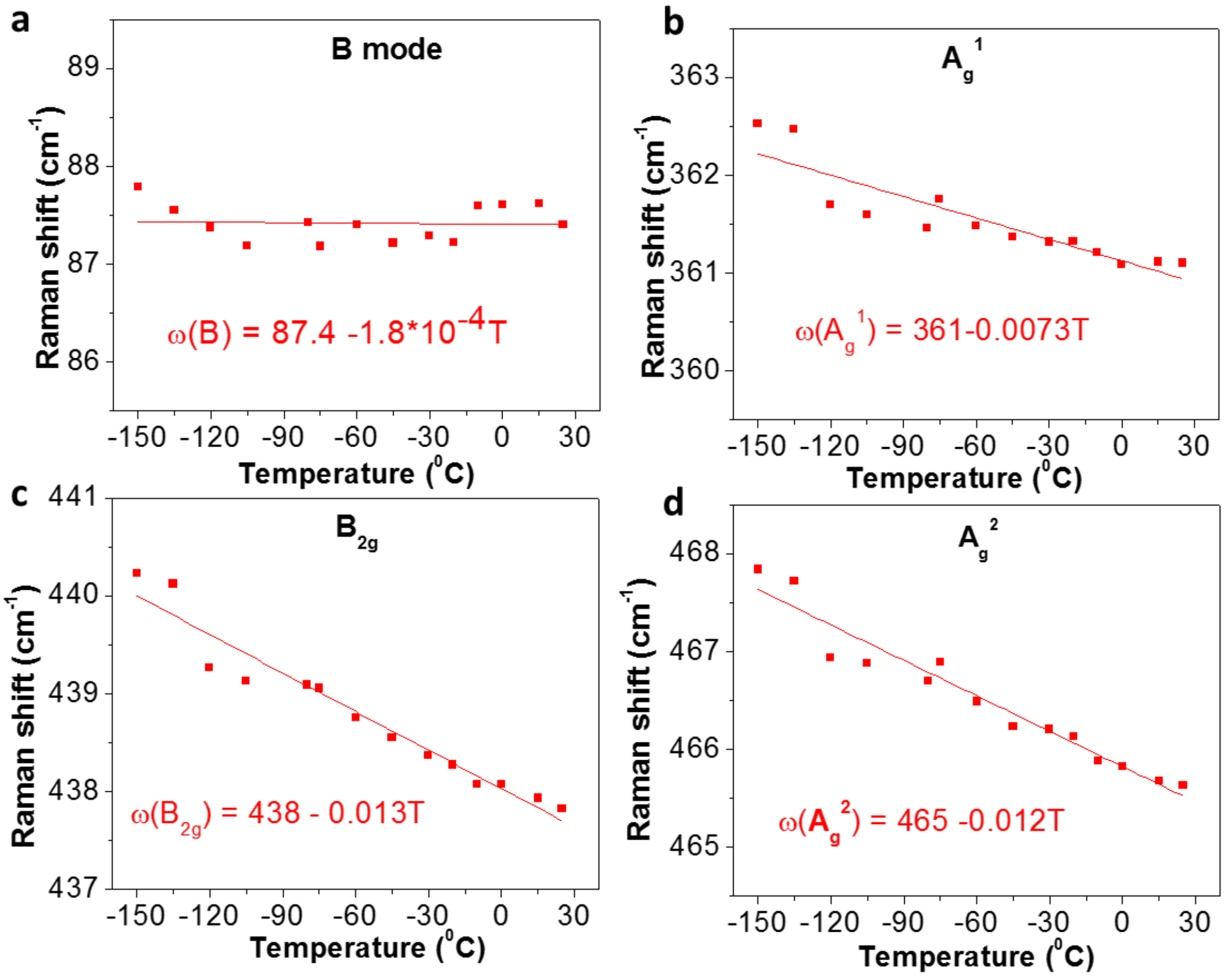

**Figure 5.** Temperature dependence of the frequency of (a) B mode, (b) $A_g^1$, (c) $B_{2g}$ and (d) $A_g^2$ modes. The red lines are the corresponding fitting lines. The laser wavelength is 633 nm.

In Fig. 5, we linearly fitted the data and considered the first-order temperature coefficient $\chi$, which has two components leading to the Raman frequency shift. In detail, the temperature dependence of the Raman frequency can be rewritten as $\omega = \omega_0 + \chi_T \Delta T + \chi_V \Delta V = \omega_0 + (\frac{\partial \omega}{\partial T})_V \Delta T + (\frac{\partial \omega}{\partial V})_T \Delta V$, where the first term $(\frac{\partial \omega}{\partial T})_V \Delta T$ is the "self-energy" shift, which is the pure temperature effect, and the second term $(\frac{\partial \omega}{\partial V})_T \Delta V$ is due to the crystal thermal expansion[64,65]. For the out-of-plane B mode of few-layer BP, which is solely due to interlayer coupling, the contribution to the Raman shift from the second term $(\frac{\partial \omega}{\partial V})_T \Delta V$ depends on the thermal expansion along the out-of-plane direction. Since the interlayer distance will not change much with the temperature, the thermal expansion along the out-of-plane direction can be ignored[69,70]. Therefore, the contribution from the crystal thermal expansion can be ignored for the B mode, as confirmed by our calculations (Fig. S6a). Thus, the harmonic behavior of the B mode in Fig. 5a suggests that the contribution from the first term ($(\frac{\partial\omega}{\partial T})_V\Delta T$) should be near zero as well, indicating the weak phonon coupling for the B mode. For the HF out-of-plane intralayer mode $A_g^1$, the contribution from the in-plane thermal expansion is also negligible (supported by the calculation results in Fig. S6b). The anharmonic behavior of $A_g^1$ mode in Fig. 5b is hence largely due to the "self-energy shift" (i.e., the anharmonic phonon coupling). While for the HF in-plane intralayer modes $B_{2g}$ and $A_g^2$, the contribution from the in-plane thermal expansion is significant (as revealed by the calculations in Figs. S6c-d), since tensile strain can be induced in the BP plane and the sequential softening of the P-P bonds occurs with increasing temperature. The anharmonic phonon effect for $B_{2g}$ and $A_g^2$ in Figs. 5c-d is thus mainly due to the decrease of the force constants by the thermal expansion (the second term) with a minor contribution from the anharmonic phonon coupling (the first term).

**Discussion**

The determination of the thickness and crystalline orientation are two crucial aspects for advancing studies of few-layer BP. Raman spectroscopy is expected to shine light on both aspects due to the non-destructive and convenient characterization. The identification of the crystalline orientation has been successfully achieved using the polarization dependence of the HF intralayer Raman modes[45,47]. However, they fail to determine the thickness of BP. In this work, for the first time, the LF interlayer breathing modes are observed in few-layer BP and show promising potential in identifying both the crystalline orientation and the thickness, as well as probing the interlayer vdW coupling. The breathing modes are assigned to the same symmetry as the HF $A_g$ modes, so they share similar properties on the laser polarization dependence. By rotating the sample under parallel polarization configuration, they all show the same intensity variation period of 180° with the strongest intensities occurring when the sample armchair direction is along the polarization direction of the light. The crucial difference is that the LF breathing modes are found to be much more sensitive to the thickness and interlayer interactions, compared to HF Raman modes. Furthermore, the temperature dependence study shows that the breathing mode has a harmonic phonon effect, while the HF modes show anharmonic phonon behaviors. These observations indicate that the phonon-phonon coupling and the electron-phonon coupling are relatively weak for the breathing mode. Our new experimental/theoretical results about phonons, especially low-frequency phonons, could be very helpful for the future studies of the electronic and thermal properties of BP thin films.

## Methods

**Sample preparation.** Few-layer BP was prepared on a 300 nm $SiO_2$/Si substrate or glass substrate by mechanical exfoliation from a bulk BP, and coated by parylene (~100 nm) or PMMA film (~300nm) immediately to avoid the degradation of BP. The locations of the flakes are identified under the optical microscope.

**Raman measurements.** The Raman spectra in Figs. 2-4 were taken under a back-scattering configuration at room temperature on a triple-grating Horiba-Yobin T64000 micro-Raman system with a 632.8 nm He-Ne laser line, 1800 lines/mm grating, a micrometer resolved XYZ scanning stage, and a ×100 objective lens of NA=0.95. The laser spot diameter is about 1 μm on the sample and the laser power is controlled at around 2.5 mW. For the polarization dependence measurement, the sample was placed on a rotation stage. The sample was rotated during the measurement every 10~15°, and the polarization of the incident light and scattered light was kept parallel. The Raman spectra in Fig. 5 for the temperature dependence study was carried out on a Horiba Jobin Yvon HR800 system with a 632.8 nm He-Ne laser line, 600 lines/mm grating, a micrometer resolved XYZ scanning stage, and a ×100 objective lens of NA=0.80. The laser power is around 1 mW on the sample. The temperature was controlled by a Linkam thermal stage THMS 600. The parameters of the Raman peaks are obtained by fitting the peaks using a Lorenzian lineshape. We chose a 632 nm (1.96 eV) excitation laser in this work instead of 532 nm (2.33 eV) laser to avoid the photolysis of BP under the high energy laser, since the bonding energy of P-P bond is around 2.1 eV.

**Theoretical methods.** Plane-wave DFT calculations were performed using the VASP package equipped with projector augmented wave (PAW) pseudopotentials for electron-ion interactions. The exchange-correlation interactions are considered in the local density approximation (LDA), as well as the generalized gradient approximation (GGA) using the Perdew-Burke-Ernzerhof (PBE) functional. For the GGA-PBE calculations, the vdW interactions between layered BP are included using the DFT-D2 approach of Grimme (denoted as PBE+D2), and the vdW density functional methods optB88-vdW (denoted as PBE+optB88) and optB86b-vdW (denoted as PBE+ optB86b). The method of PBE+optB88 is used for the systematic study. For bulk BP, both atoms and cell volume were allowed to relax until the residual forces were below 0.001 eV/Å, with a cutoff energy set at 500 eV and a 12×4×9 k-point sampling in the Monkhorst-Pack scheme. By taking the in-plane zigzag direction as the X axis, the out-of-plane direction as the Y axis, and in-plane armchair direction as the Z axis, the optimized lattice parameters of bulk BP are a=3.35 Å, b=10.67 Å and c= 4.45 Å using optB88-vdW. Single- and few-layer BP systems were then modeled by a periodic slab geometry using the optimized in-plane lattice constant of the bulk. A vacuum region of 22 Å in the direction normal to the plane (Y direction) was used to avoid spurious interactions with replicas. For the 2D slab calculations, all atoms were relaxed until the residual forces were below 0.001 eV/Å and 12×1×9 k-point samplings were used (see more details and references about theoretical methods, especially the non-resonant Raman calculations, in Supporting Information).

**Acknowledgements**

The authors thank Prof. Fengnian Xia, Prof. Han Wang and Sangyeop Lee for their useful discussion and help. X.L., S.H. and M.S.D. at MIT acknowledge grant NSF/DMR-1004147 and DE-SC0001299 for financial support. Part of the Raman measurements was conducted at the Center for Nanophase Materials Sciences, which is sponsored at Oak Ridge National Laboratory by the Scientific User Facilities Division, Office of Basic Energy Sciences, U.S. Department of Energy. The theoretical work at Rensselaer Polytechnic Institute (RPI) was supported by New York State under NYSTAR program C080117 and the Office of Naval Research. The computations were performed using the resources of the Center for Computational Innovation at RPI.

**Author contributions**

X.L., S.H. J.K. and M.S.D. conceived the research. X.L., S.H. and A.A.P. performed Raman measurements and analyzed the data. X.L. and S.H. carried out the rest of experimental measurements. L.L. and V.M. performed the theoretical analysis. X.L., L.L., S.H., V.M. and M.S.D. wrote the paper. All the authors discussed the results and commented on the manuscript.

# Supplementary Information

## Observation of Low-frequency Interlayer Breathing Modes in Few-layer Black Phosphorus

*Xi Ling*[1,]*, *Liangbo Liang*[2,]*, *Shengxi Huang*[1], *Alexander A. Puretzky*[3], *David B. Geohegan*[3], *Bobby G. Sumpter*[3,4], *Jing Kong*[1], *Vincent Meunier*[2], *Mildred S. Dresselhaus*[1,5]

---

[1]Department of Electrical Engineering and Computer Science, Massachusetts Institute of Technology, Cambridge, Massachusetts 02139, USA. [2]Department of Physics, Applied Physics, and Astronomy, Rensselaer Polytechnic Institute, Troy, New York 12180, USA. [3]Center for Nanophase Materials Sciences, Oak Ridge National Laboratory, Oak Ridge, Tennessee 37831, USA. [4]Computer Science and Mathematics Division, Oak Ridge National Laboratory, Oak Ridge, Tennessee 37831, USA. [5]Department of Physics, Massachusetts Institute of Technology, Cambridge, Massachusetts 02139, USA. *These authors contributed equally to this work. Correspondence should be addressed to M.S.D. (email: mdress@mit.edu), V.M. (email: meuniv@rpi.edu) and X.L. (email: xiling@mit.edu).

**Section S1. Thickness-dependent interlayer breathing modes in BP**

**Table S1.** Calculated frequencies (in $cm^{-1}$) of interlayer breathing (B) modes of 2L and bulk BP using different theoretical methods. The experimental frequency of the bulk B mode obtained using the inelastic neutron scattering is also listed for comparison[48–50,54]. Clearly, for the bulk, PBE+optB88 yields the best match to the experimental frequency. PBE significantly underestimates the value, while other methods LDA, PBE+D2 and PBE+optB86b overestimate. Thus, the method of PBE+optB88 is adopted for the systematic study. More details about the theoretical methods are in Section S6.

| Methods | LDA | PBE | PBE+D2 | PBE+optB86b | PBE+optB88 | Experiment |
|---|---|---|---|---|---|---|
| 2L | 78.6 | 33.6 | 73.8 | 66.3 | 62.7 | |
| bulk | 105.5 | 46.6 | 103.3 | 91.3 | 86.1 | 87.1 |

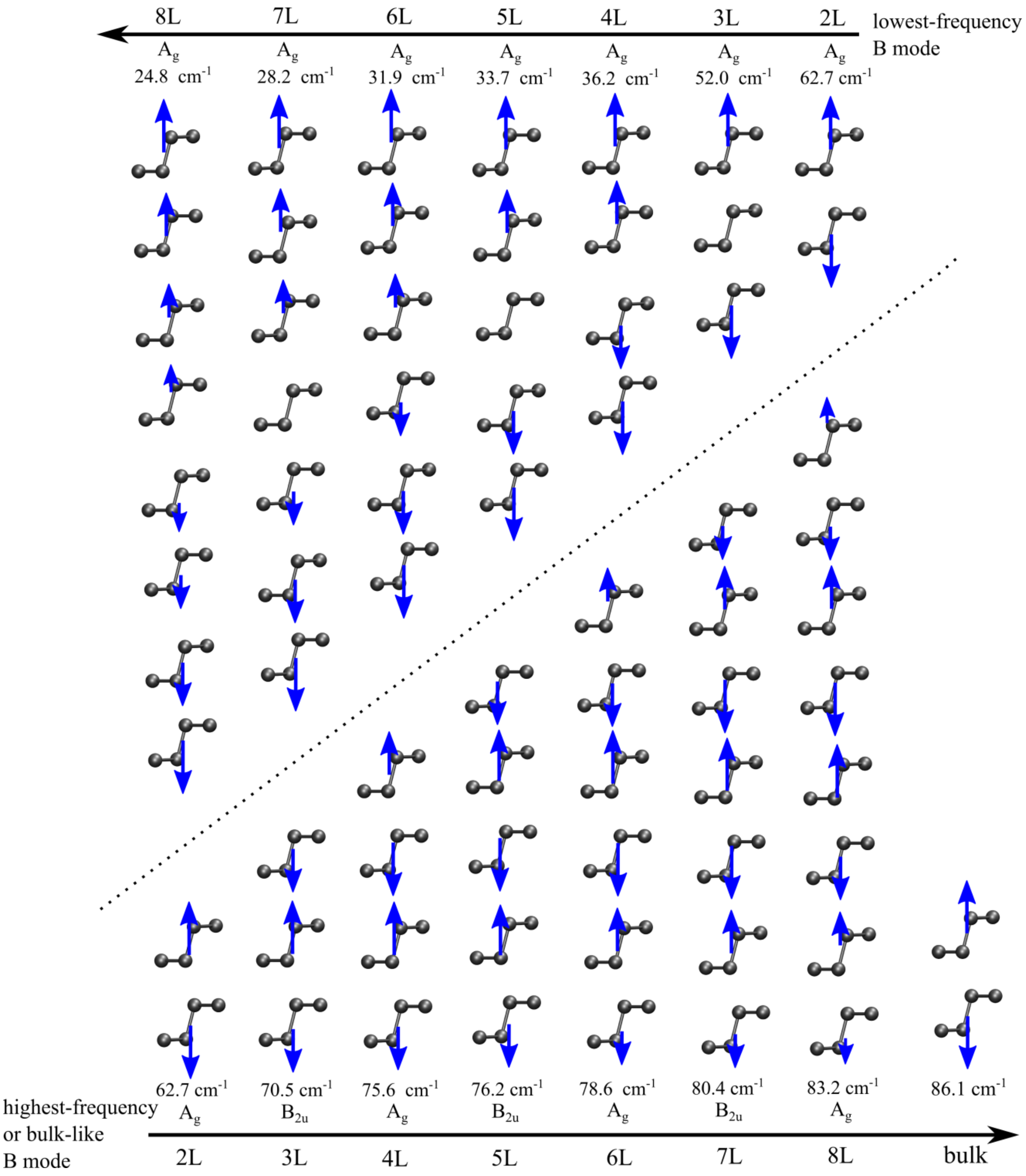

**Figure S1. Calculated vibrations and frequencies of the highest-frequency and lowest-frequency B modes for 2L to 8L BP using PBE+optB88.** Each blue arrow indicates the displacement of a whole layer. The highest-frequency B mode of any thickness is the bulk-like B mode, where each adjacent layer vibrates in the opposite directions like the bulk. With increasing thickness, it blue shifts and approaches the bulk limit ~87 cm$^{-1}$. For the lowest-frequency B mode, generally, the top or bottom half segment shows in-phase displacements, and it is the two segments that vibrate in the opposite directions. Compared to the highest-frequency one, a greater proportion of in-phase displacements in the lowest-frequency B mode lead to lower frequencies. Note that for 2L, there is only one B mode (thus it is both the highest-frequency and lowest-frequency B mode).

**Section S2. Polarization dependence of Raman-active modes in BP**

As mentioned in the main text, by denoting X axis as the sample in-plane zigzag direction, Y axis as the out-of-plane direction, and Z axis as the in-plane armchair direction, the Raman tensors $\tilde{R}$ of Raman-active modes $A_g$, $B_{1g}$, $B_{2g}$ and $B_{3g}$ are

$$\tilde{R}(A_g)=\begin{pmatrix} a & \cdot & \cdot \\ \cdot & b & \cdot \\ \cdot & \cdot & c \end{pmatrix}, \qquad \tilde{R}(B_{1g})=\begin{pmatrix} \cdot & d & \cdot \\ d & \cdot & \cdot \\ \cdot & \cdot & \cdot \end{pmatrix},$$

$$\tilde{R}(B_{2g})=\begin{pmatrix} \cdot & \cdot & e \\ \cdot & \cdot & \cdot \\ e & \cdot & \cdot \end{pmatrix}, \qquad \tilde{R}(B_{3g})=\begin{pmatrix} \cdot & \cdot & \cdot \\ \cdot & \cdot & f \\ \cdot & f & \cdot \end{pmatrix}. \qquad \text{(S1)}$$

In the typical experimental back-scattering laser geometry (Y in and Y out), the electric polarization vectors of the incident and scattered light $e_i$ and $e_s$ are in-plane (the X-Z plane). By setting the polarization angle of the incident (scattered) light as $\theta$ ($\gamma$) with respect to X axis, $e_i$ = $(cos\theta, 0, sin\theta)$ and $e_s$ =$(cos\gamma, 0, sin\gamma)$, Since Raman intensity $I \propto |e_i \cdot \tilde{R} \cdot e_s^T|^2$, we then have

$$I \propto \left| (cos\theta, 0, sin\theta) \quad \tilde{R} \quad \begin{pmatrix} cos\gamma \\ 0 \\ sin\gamma \end{pmatrix} \right|^2. \qquad \text{(S2)}$$

Applying the Raman tensors $\tilde{R}$ in Eq. S1 to Eq. S2, we can obtain

$$I_{A_g} \propto a^2 \left| cos\theta cos\gamma + \frac{c}{a} sin\theta sin\gamma \right|^2, \; I_{B_{2g}} \propto e^2 sin^2(\theta + \gamma),$$

$$I_{B_{1g}} = 0, \qquad I_{B_{3g}} = 0. \qquad \text{(S3)}$$

Therefore, $B_{1g}$ and $B_{3g}$ cannot be observed, while only $A_g$ and $B_{2g}$ modes can be observed.

In general, there are two methods to study the polarization dependence. One is by rotating the sample while fixing the polarization of the incident and scattered light; the other is by changing the polarization of the incident or scattered light while fixing the sample. In this work, we have used the first method. The polarization angle of the incident and scattered light $\theta$ and $\gamma$ are fixed, and the sample is rotated in-plane (the X-Z plane) by $\varphi$ with respect to X axis. The rotation matrix and its inverse are

$$r=\begin{pmatrix} cos\varphi & 0 & -sin\varphi \\ 0 & 1 & 0 \\ sin\varphi & 0 & cos\varphi \end{pmatrix} \text{ and } r^{-1}=\begin{pmatrix} cos\varphi & 0 & sin\varphi \\ 0 & 1 & 0 \\ -sin\varphi & 0 & cos\varphi \end{pmatrix}. \qquad \text{(S4)}$$

Consequently, for any Raman tensor $\tilde{R}$, the intensity becomes

$$I \propto \left|(cos\theta, 0, sin\theta)\ r \quad \tilde{R} \quad r^{-1} \quad \begin{pmatrix} cos\gamma \\ 0 \\ sin\gamma \end{pmatrix}\right|^2$$

$$\propto \left|(cos\theta, 0, sin\theta)\ \begin{pmatrix} cos\varphi & 0 & -sin\varphi \\ 0 & 1 & 0 \\ sin\varphi & 0 & cos\varphi \end{pmatrix} \quad \tilde{R} \quad \begin{pmatrix} cos\varphi & 0 & sin\varphi \\ 0 & 1 & 0 \\ -sin\varphi & 0 & cos\varphi \end{pmatrix}\ \begin{pmatrix} cos\gamma \\ 0 \\ sin\gamma \end{pmatrix}\right|^2$$

$$\propto \left|(cos\theta cos\varphi + sin\theta sin\varphi, 0, -cos\theta sin\varphi + sin\theta cos\varphi) \quad \tilde{R} \quad \begin{pmatrix} cos\varphi cos\gamma + sin\varphi sin\gamma \\ 0 \\ -sin\varphi cos\gamma + cos\varphi sin\gamma \end{pmatrix}\right|^2$$

$$\propto \left|(\cos(\theta - \varphi), 0, \sin(\theta - \varphi)) \quad \tilde{R} \quad \begin{pmatrix} \cos(\gamma - \varphi) \\ 0 \\ \sin(\gamma - \varphi) \end{pmatrix}\right|^2. \tag{S5}$$

Compared to Eq. S2, we can infer that the rotation of the crystal sample by $\varphi$ is equivalent to rotation of the laser polarization of both incident and scattered light by $-\varphi$ with the sample fixed. Under the parallel polarization configuration ($\gamma = \theta$), based on Eq. S3 and Eq. S5, we then have

$$I_{A_g} \propto a^2 \left|cos^2(\varphi - \theta) + \frac{c}{a} sin^2(\varphi - \theta)\right|^2 \propto a^2 \left|1 + (\frac{c}{a} - 1) sin^2(\varphi - \theta)\right|^2,$$

$$I_{B_{2g}} \propto e^2 sin^2 2(\varphi - \theta),$$

$$I_{B_{1g}} = 0, I_{B_{3g}} = 0. \tag{S6}$$

Since $\theta$ is fixed, the intensity of an $A_g$ mode depends on both the sample rotation angle $\varphi$ and the ratio *c*/*a*, while the intensity of a $B_{2g}$ mode only depends on the rotation angle $\varphi$. Clearly, the intensity variation period is always 90° for a $B_{2g}$ mode: the intensity reaches the minimum at $\varphi = \theta$ or $\theta + 90°$ or $\theta + 180°$ or $\theta + 270°$, and the maximum at $\varphi = \theta + 45°$ or $\theta + 135°$ or $\theta + 225°$ or $\theta + 315°$, as shown in Figure S2b. For an $A_g$ mode, the situation is more complicated due to the ratio *c*/*a*. If *c*/*a* = 1 (i.e., isotropic), then $I_{A_g} \propto a^2$ always, which has no dependence on the polarization (see Fig. S2a). However, for anisotropic BP, *c*/*a* ≠ 1. If *c/a* > 1 (Fig. S2a), the intensity variation period of an $A_g$ mode is 180°, with the minimum intensity $I_{A_g} \propto a^2$ at $\varphi = \theta$ or $\theta + 180°$ (the sample is rotated by $\theta$ and now zigzag direction is along the polarization direction of incident light), and the maximum intensity $I_{A_g} \propto c^2$ at $\varphi = \theta + 90°$ or $\theta + 270°$ (the sample is rotated by $\theta + 90°$ and now armchair direction is along the polarization direction of incident light). If $c/a < 1$, the intensity variation period is still 180°, but the minimum and maximum intensity angles switch. According to our calculations for 1L-4L BP and a previous experimental work for few-layer BP[45], *c* (Raman tensor component in the armchair direction) is expected to be larger than *a* (Raman tensor component in the zigzag direction), hence $c/a > 1$. In short, by rotating the crystal sample under parallel polarization configuration, the intensity variation period is always 180° for an $A_g$ mode, while it is 90° for a $B_{2g}$ mode. Additionally, when the sample armchair (zigzag) direction is along the polarization direction of incident light, an $A_g$ mode shows the maximum (minimum) intensity, while a $B_{2g}$ mode is forbidden, as illustrated by the calculated polarization dependence of an $A_g$ mode ($c/a$ = 1, 2, 4, 8) and an $B_{2g}$ mode in Fig. S2. In our calculations, the polarization angle of the incident light $\theta$ is set at 0° so that the initial zigzag direction is along the polarization direction of incident light before sample rotation. Thus, the minimum and maximum intensity rotation angles of an $A_g$ mode are always 0° and 90° respectively, despite of different ratios ($c/a$ = 2, 4, 8) in Fig. S2a. What is different is the quickly decreased minimum/maximum intensity ratio $I_{min}/I_{max} \propto (a/c)^2$. As a result, when $c/a$ = 2, an $A_g$ mode still shows relatively strong intensities at the minimum intensity rotation angle; when $c/a$ = 8, an $A_g$ mode is almost forbidden at the minimum intensity rotation angle. These results can well explain the observed minor differences between polarization dependence of LF breathing modes and HF $A_g^1$ and $A_g^2$ modes in Figs. 3b-g of the main text.

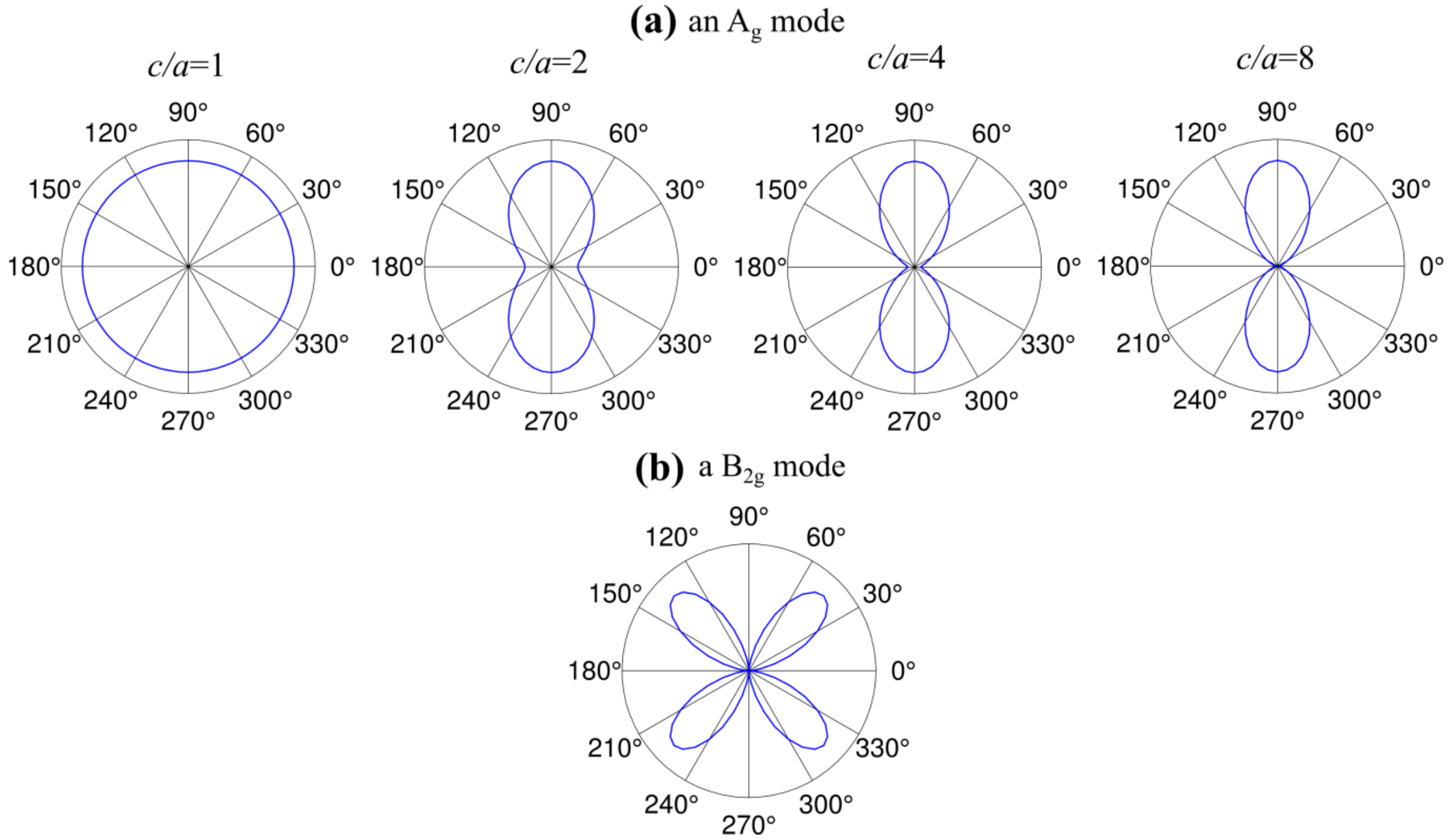

**Figure S2. Theoretical polarization dependence by rotating the sample.** Polar plots of calculated Raman intensities of (a) $A_g$ and (b) $B_{2g}$ modes as a function of crystal rotation angle $\varphi$. The polarization angle of the incident light $\theta$ is set at 0° so that the sample initial zigzag direction is along the polarization direction of incident light. In (a), different *c*/*a* ratios are considered.

Now we show how to determine the crystalline orientation of the BP flake based on its polarization dependence measured in Fig. 3. According to Figs. 3b-g, after ~45° clockwise rotation of the BP sample, the intensities of $A_g$ modes reach the maximum and thus the sample armchair direction is now along the polarization direction of incident light (i.e., the horizontal direction in the image of Fig. 2a). Therefore before the rotation, the armchair direction of the sample is around 45° of the image in Fig. 2a.

Note that in the experimental back-scattering geometry, according to Eq. S3 and Eq. S6, the intensities of $B_{1g}$ and $B_{3g}$ modes are zero under both methods, and thus their polarization dependence cannot be probed. If the BP sample is titled out-of-plane instead of rotated in-plane, they could be observed. From Eq. S1, Raman tensors of $B_{1g}$ and $B_{3g}$ modes share similar formats to that of the $B_{2g}$ mode (i.e., non-zero terms are all off-diagonal), indicating that their polarization dependence should be similar to the $B_{2g}$ mode.

**Section S3. High-frequency Raman spectra at different crystal rotation angles**

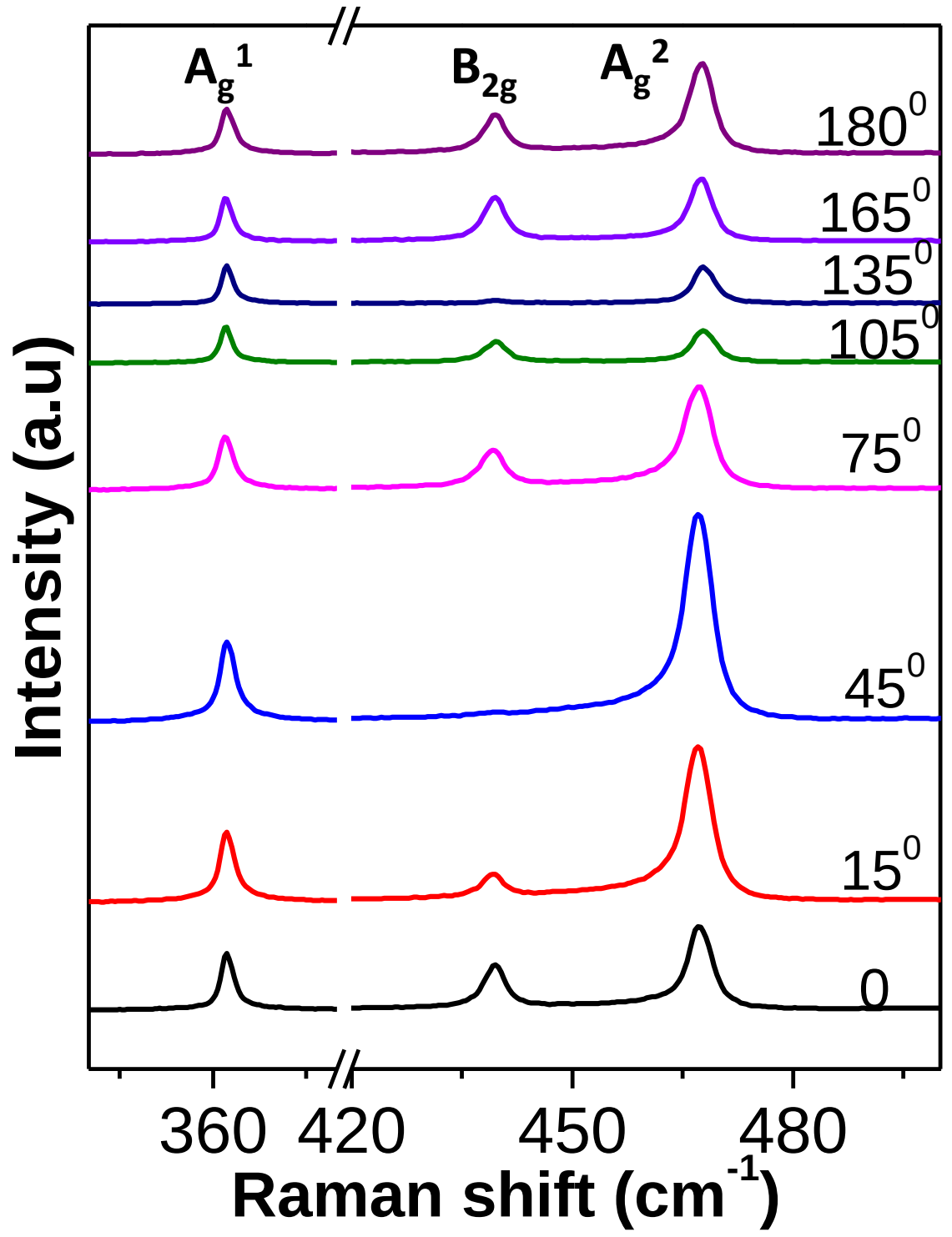

**Figure S3.** Raman spectra in the high-frequency region of the few-layer BP (corresponding to Fig. 2a) at the different rotation angles.

**Section S4. Thickness dependence of the high-frequency Raman modes**

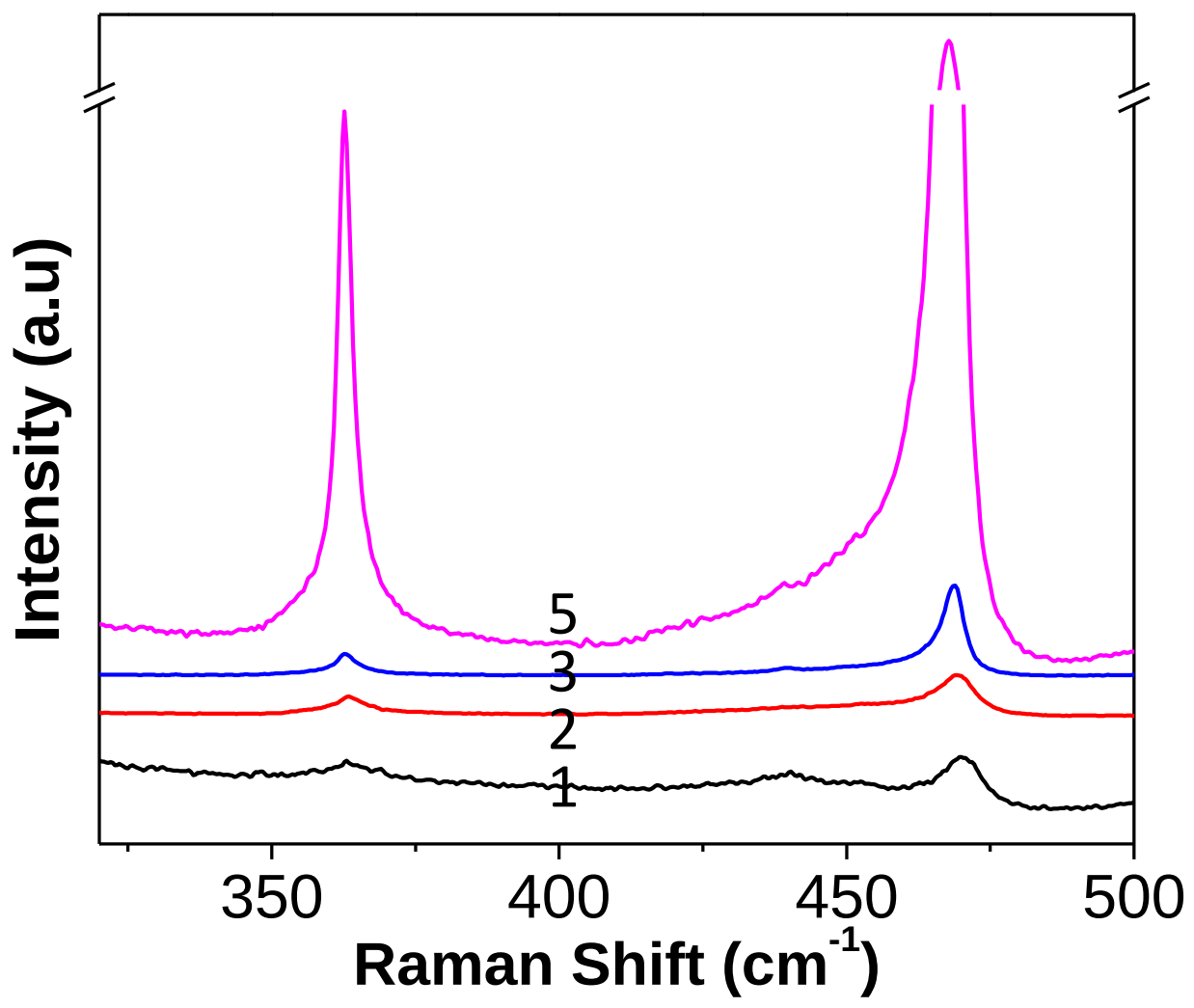

**Figure S4.** High-frequency Raman spectra of different BP flakes. The labels are corresponding to those in Fig. 4.

## Section S5. Temperature dependence of the Raman modes

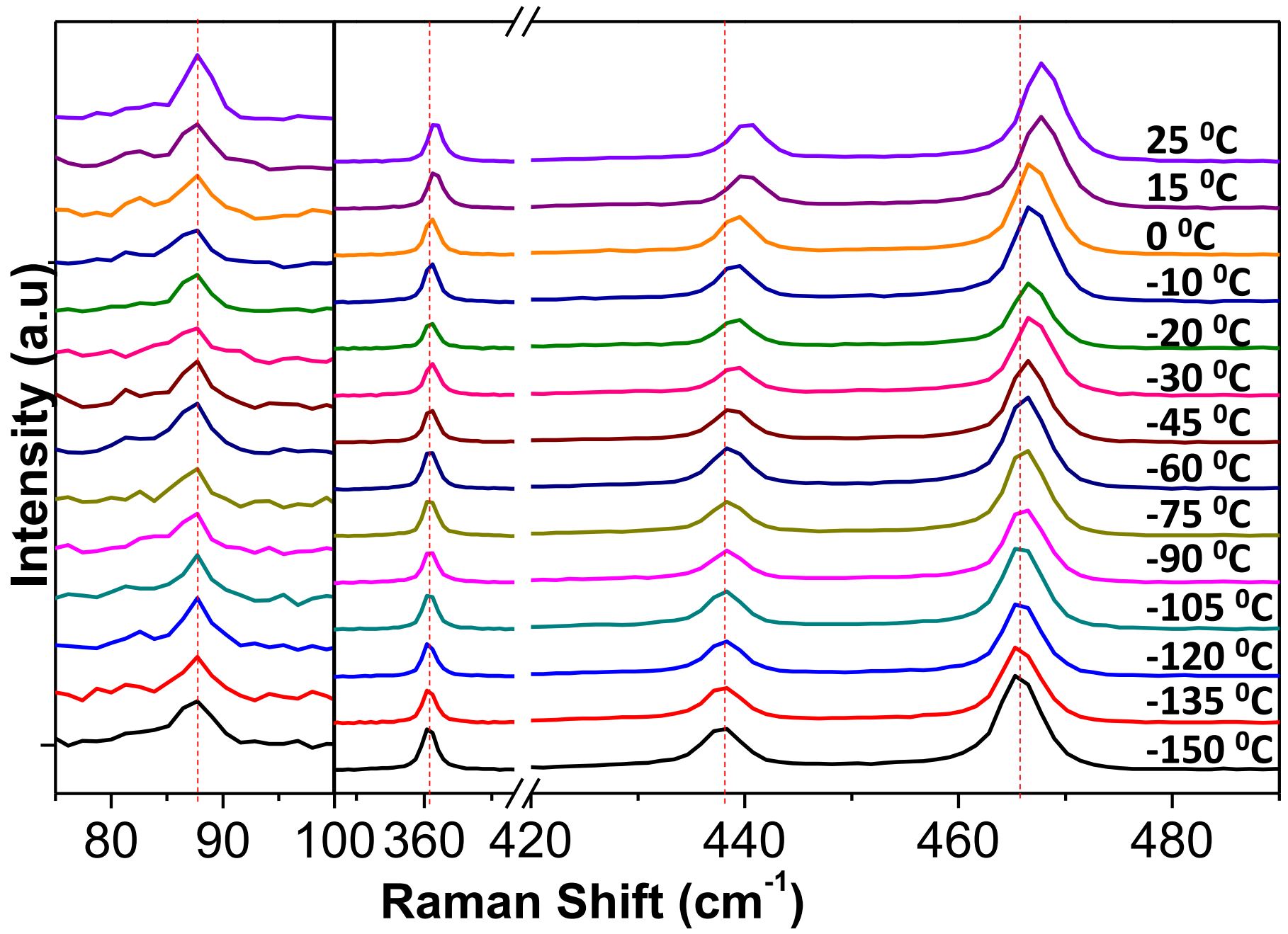

**Figure S5.** Raman spectra of few-layer BP at different temperatures.

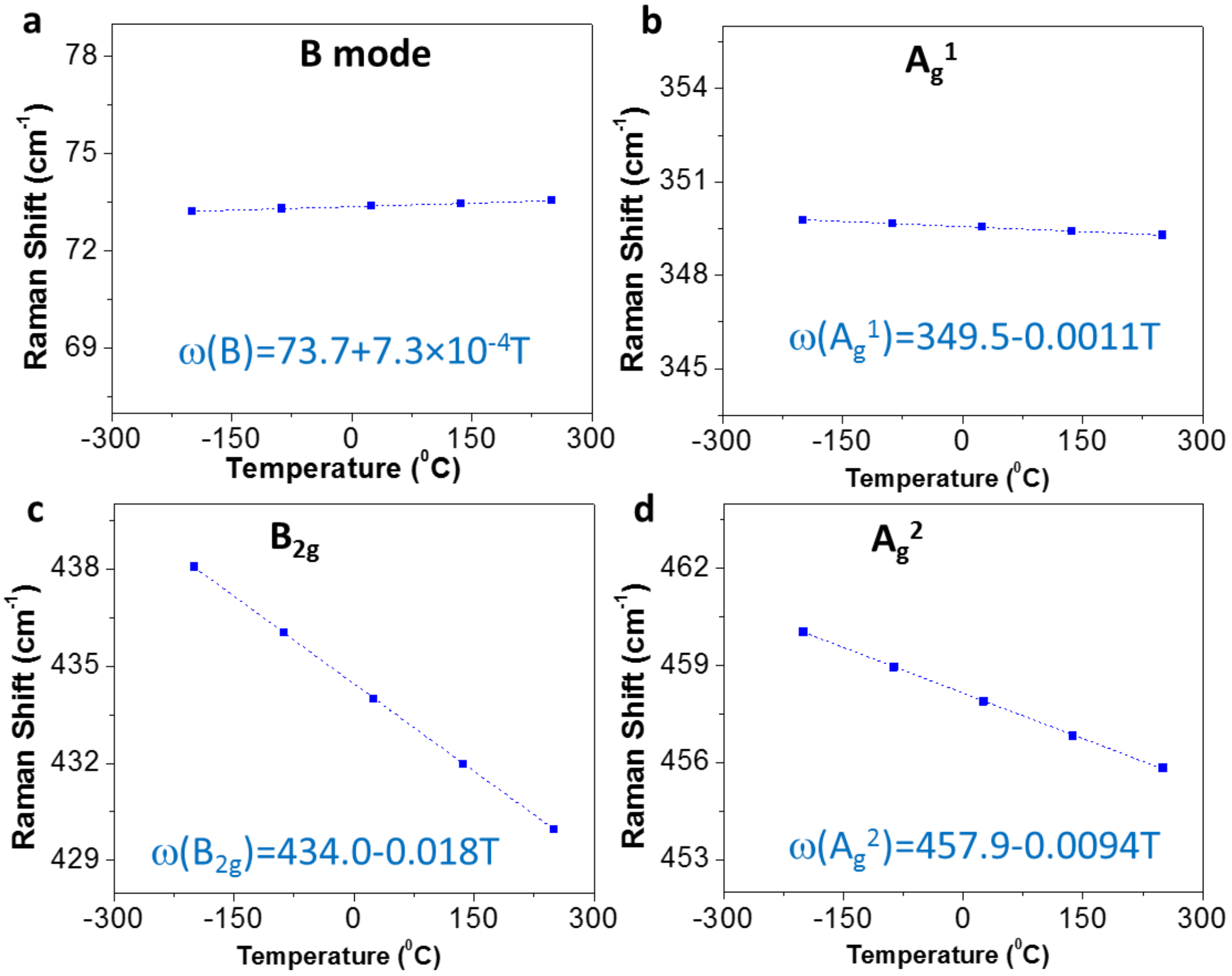

**Figure S6.** Calculated temperature dependence of the frequencies of (a) B mode, (b) $A_g^1$, (c) $B_{2g}$ and (d) $A_g^2$ modes considering the contribution of the thermal expansion (the second term).

## Section S6. Theoretical methods

Plane-wave DFT calculations were performed using the VASP package equipped with projector augmented wave (PAW) pseudopotentials for electron-ion interactions[71,72]. Previous theoretical calculations have demonstrated that the geometrical and electronic properties of bulk and few-layer BP are highly functional dependent[25,73]. Therefore, for comparison and completeness purpose in this work, the exchange-correlation interactions are considered in the local density approximation (LDA), as well as the generalized gradient approximation (GGA) using the Perdew-Burke-Ernzerhof (PBE) functional[74]. For the GGA-PBE calculations, the vdW interactions between layered BP are included using the DFT-D2 approach of Grimme (denoted as PBE+D2)[75], and the vdW density functional methods optB88-vdW (denoted as PBE+optB88) and optB86b-vdW (denoted as PBE+ optB86b)[76]. For bulk BP, both atoms and cell volume were allowed to relax until the residual forces were below 0.001 eV/Å, with a cutoff energy set at 500 eV and a 12×4×9 k-point sampling in the Monkhorst-Pack scheme[77]. By taking the in-plane zigzag direction as the X axis, the out-of-plane direction as the Y axis, and in-plane armchair direction as the Z axis, the optimized lattice parameters of bulk BP are a=3.35 Å, b=10.67 Å and c= 4.45 Å using optB88-vdW. Single- and few-layer BP systems were then modeled by a periodic slab geometry using the optimized in-plane lattice constant of the bulk. A vacuum region of 22 Å in the direction normal to the plane (Y direction) was used to avoid spurious interactions with replicas. For the 2D slab calculations, all atoms were relaxed until the residual forces were below 0.001 eV/Å and 12×1×9 k-point samplings were used.

Then non-resonant Raman calculations were performed using the fully relaxed geometries. Since Raman intensity $I \propto |e_i \cdot \tilde{R} \cdot e_s^T|^2$, the calculations of Raman tensors $\tilde{R}$ are of most importance, which require the information of phonon frequencies, phonon eigenvectors (i.e., vibrations) and the changes of the polarizability or dielectric constant tensors with respect to phonon vibrations (see more details and equations in Ref.55 )To obtain Raman scattering, one needs to calculate the dynamic matrix and derivatives of the dielectric constant tensors. The dynamic matrix was calculated using the finite difference scheme, implemented in the Phonopy software[52,78]. Hellmann-Feynman forces in the 3×1×3 supercell were computed by VASP for both positive and negative atomic displacements (δ = 0.03 Å) and then used in Phonopy to construct the dynamic matrix, whose diagonalization provides phonon frequencies and eigenvectors. Phonopy was also used to determine the space and point groups of a system, and the symmetry of each phonon mode. The derivatives of the dielectric constant tensors were also calculated by the finite difference approach. For both positive and negative atomic displacements in the single unit cell, the dielectric constant tensors were computed by VASP using density functional perturbation theory and then their derivatives can be obtained. With phonon frequencies, phonon eigenvectors and derivatives of the dielectric constant tensors, Raman tensors $\tilde{R}$ can be computed. Then Raman intensity for every phonon mode was obtained for a given laser polarization set-up to finally yield a Raman spectrum after Gaussian broadening.